\documentclass[11pt,oneside,reqno]{amsart}

\usepackage{amsmath,amssymb}
\usepackage{amsthm,amsfonts}
\usepackage{amssymb,latexsym}
\usepackage{graphicx,rotating}
\usepackage{exscale} 
\usepackage{upref} 
\usepackage{txfonts}
\usepackage{hyperref}

\theoremstyle{plain}

\theoremstyle{definition}

\theoremstyle{remark}

\numberwithin{equation}{section}

\def\E{{\mathbb E}}

\def\R{{\mathbb R}}

\begin{document}

\title[Inference for the Birnbaum--Saunders distribution]{On Birnbaum--Saunders 
inference}

\author{Audrey H.M.A.\ Cysneiros}
\author{Francisco Cribari--Neto}
\author{Carlos A.G.\ Ara\'ujo Jr}
\address{Departamento de Estat\'{\i}stica, CCEN\\
         Universidade Federal de Pernambuco\\
     Cidade Universit\'{a}ria\\
         Recife/PE, 50740--540, Brazil}
\email[F.\ Cribari--Neto]{cribari@de.ufpe.br}
\email[A.H.M.A.\ Cysneiros]{audrey@de.ufpe.br}
\email[C.A.G. Ara\'ujo Jr]{carlosgadelha@cox.de.ufpe.br}
\urladdr[F.\ Cribari--Neto]{\url{http://www.de.ufpe.br/~cribari}}

%\thanks{Research supported by CNPq and CAPES. We also thank two 
%anonymous referees for their suggestions.}


\date{\today}

\begin{abstract}
The Birnbaum--Saunders distribution, also known as 
the fatigue-life distribution, is frequently used in reliability studies. 
We obtain adjustments to the Birnbaum--Saunders profile likelihood function.
The modified versions of the likelihood function were obtained for both 
the shape and scale parameters, i.e., we take the shape parameter to be 
of interest and the scale parameter to be of nuisance, and then consider 
the situation in which the interest lies in performing inference on the 
scale parameter with the shape parameter entering the modeling in 
nuisance fashion. Modified profile maximum likelihood 
estimators are obtained by maximizing the corresponding adjusted likelihood 
functions. We present numerical evidence on the finite sample behavior of 
the different estimators and associated likelihood ratio tests. The results 
favor the adjusted estimators and tests we propose. 
A novel aspect of the profile likelihood adjustments obtained in 
this paper is that they yield improved point estimators {\em and\/} 
tests.  The two profile 
likelihood adjustments work well when inference is made on the shape 
parameter, and one of them displays superior behavior when it comes to 
performing hypothesis testing inference on the scale parameter. 
Two empirical applications are briefly presented. 
\end{abstract}

\maketitle

\section{Introduction}\label{S:introduction}

Birnbaum and Saunders~(1969a) derived a two-parameter distribution
using a set-up in which failure time due to fatigue under cyclic loading 
when failure follows from the development and growth 
of a dominant crack. 
According to Marshall and Olkin~(2007), the Birnbaum--Saunders 
distribution has appeared in several different contexts, and various 
derivations of the distribution are known. According to them (pp.\ 466-467),
``it was given by Fletcher~(1911), and according to Schr\"odinger~(1915)
it was obtained by Konstantinowsky~(1914);'' additionally, 
``it was obtained by Freudental and Shinozuka~(1961), but it was 
the derivation of Birnbaum and Saunders~(1969a) that brought the 
usefulness of the distribution into clear focus.'' 
Desmond~(1985) derived the same distribution in a 
more general setting; he used a biological model and relaxed several of the 
assumptions made by the original authors. The relationship between the 
the Birnbaum--Saunders and inverse Gaussian distributions was explored by 
Desmond~(1986). It can shown that the Birnbaum--Saunders distribution 
is a mixture between an inverse Gaussian distribution and a generalized 
inverse Gaussian distribution; see Bhattacharyya and Fries~(1982).

The random variable $T$ is said to be Birnbaum--Saunders distributed, denoted   
$T \sim \mathcal{BS}(\alpha,\beta)$, if its density function is given by 
$$\small{
f(t;\alpha,\beta)={1\over2\sqrt{2\pi}\alpha\beta}\left[\left({\beta\over\ 
t}\right)^{1/2}
+ \left({\beta\over t}\right)^{3/2}\right]\exp\left[-{1\over 
2\alpha^2}\left({t\over\beta} +
{\beta\over t} - 2\right)\right],}
$$
$t, \alpha, \beta > 0$,
where $\alpha$ is the shape parameter and $\beta$ is the scale parameter.  
It is noteworthy that the reciprocal property holds for the Birnbaum--Saunders 
distribution: $T^{-1} \sim \mathcal{BS}(\alpha, \beta^{-1})$; see 
Saunders~(1974). It is easy to show that 
$\E(T) = \beta\left(1+\frac{1}{2}\alpha^2\right)$, 
$\mathrm{var}(T) = (\alpha \beta)^2\left(1+\frac{5}{4}\alpha^2\right)$,
$\E(T^{-1}) = \beta^{-1}\left(1+\frac{1}{2}\alpha^2\right)$
and
$\mathrm{var}(T^{-1}) = \alpha^2\beta^{-2}\left(1+\frac{5}{4}\alpha^2\right)$.

The Birnbaum--Saunders distribution function is 
$$F(t;\alpha,\beta)=\Phi\left({1\over\alpha}\left[\left({t\over\beta}\right)^{1/2}
- \left({\beta\over t}\right)^{1/2}\right]\right),
\quad 0 < t < \infty,\ \ \alpha,\beta > 0,$$
where $\Phi(\cdot)$ denotes the standard normal distribution function. 
Note that $\beta$ is the median of the distribution: 
$F_{T}(\beta) = \Phi(0) = 0.5$.
%Mann, Schafer and Singpurwalla (1974, p.\ 155) noted that, although the 
%hazard rate implied by the Birnbaum--Saunders distribution is not increasing, 
%the average hazard rate is nearly nondecreasing. 
It was shown by 
Kundu, Kannan and Balakishnan~(2008) that the Birnbaum--Saunders 
hazard function is an upside down function for all values of the shape 
and scale parameters. Hence, the distribution is useful in a number 
of practical situations where the hazard function increases up to a 
point and then decreases. The authors have also addressed the important issue 
of performing inference on the point at which the hazard function 
reaches its maximum. 

Oftentimes the interest lies in performing inference on a subset of the parameters 
that index the model; such parameters are said to be of {\em interest}, and the 
remaining ones are {\em nuisance parameters}. 
For instance, in Birnbaum--Saunders reliability studies, one is typically 
interested in performing inference on one of the parameters that index 
the model, the other parameter entering the modeling process in nuisance 
fashion. In the presence of nuisance parameters, inferences 
are usually based on the profile likelihood function, which is obtained by 
replacing, in the likelihood function, the nuisances parameters by their 
corresponding maximum likelihood estimators for fixed values of the parameters 
of interest. 
The resulting function --- the profile likelihood function --- 
will only depend on the parameters of interest. It is noteworthy, however, that 
such a function is not a true likelihood function, and some of the properties 
that hold for likelihood functions are no longer valid; in particular, there 
exist score and information biases that do not vanish as the sample size 
increases. Several adjustments to the profile likelihood function have been 
proposed; see, e.g., Barndorff--Nielsen (1983, 1994), Cox and Reid (1987, 
1992), McCullagh and Tibshirani (1990) and Stern (1997). The main idea 
behind these adjustments is to 
add a term to the log-likelihood function prior to maximizing it, in order 
to overcome the aforementioned shortcomings. 

In this paper we shall use the results in Barndorff--Nielsen (1983), 
Severini (1998, 1999) and Cox and Reid (1987) to obtain adjustments to the 
Birnbaum--Saunders profile likelihood function. A novel aspect of this 
approach is that the effect of the nuisance parameter on the inference 
performed  
on the interest parameter is greatly reduced. It is also noteworthy that 
likelihood ratio type tests constructed using the adjusted profile likelihood 
function typically have superior finite sample performance. In short, by 
adjusting the profile likelihood function and then maximizing it one can 
perform reliable point estimation and hypothesis testing inference even when 
the sample size is small. Our results shall allow practitioners to perform 
reliable inference when using the Birnbaum--Sanders model in small samples. 
A motivation for our analysis lies in the important situation in which 
one wishes to make inferences on the the median failure time in a 
reliability study. As we have seen, the median of the Birnbaum--Saunders
distribution is $\beta$, one of the parameters that index such a 
distribution. Therefore, here  $\alpha$ is a nuisance parameter.  
There are also situations where the interest lies in 
performing statistical inference on the shape parameter $\alpha$, 
with $\beta$ figuring as a nuisance unknown quantity.  
It is thus important to develop reliable and accurate inference 
strategies that are not sensitive (or, at least, less sensitive) 
to the parameter that enters the 
modeling in nuisance fashion. This is our chief goal. 

The paper unfolds as follows. 
Section \ref{S:likelihoods} introduces adjustments to the profile likelihood 
function when the interest lies in performing inference in the presence of 
nuisance parameters. 
In Section \ref{S:BSlikelihoods}, we derive adjustments to the 
Birnbaum--Saunders profile likelihood function. The use of such adjustments 
delivers, as noted above, improved estimation \textit{and} testing inference in small samples. 
Alternative inference strategies are presented in Section \ref{S:alternativeestimators}. 
Numerical results are presented in Sections \ref{S:MonteCarlo1} and 
\ref{S:MonteCarlo2}, and two applications are presented in Section 
\ref{S:applications}. Finally, Section \ref{S:conclusions} summarizes 
the main findings and lists directions for future research.

\section{Profile likelihood function and adjustments}\label{S:likelihoods}

Let $t_{1},\ldots,t_{n}$ be independent and identically distributed random 
variables with joint density $f(t;\theta)$, where $\theta \subseteq \R^p$ 
%I\!\!R^p$ 
is a $p$-vector of unknown parameters and $\mathbf{t}=(t_{1},\ldots,t_{n})^{\top}$. 
In what follows, we shall partition $\theta$ as $\theta=(\tau^{\top},\phi^{\top})^{\top}$, 
where $\tau$, a $q$-vector, contains the parameters of interest and $\phi$, a 
$(p-q)$-vector, contains the nuisance parameters. 

Inference can be based on the profile likelihood function, defined as 
$L_p(\tau)=L(\tau,\widehat{\phi}_\tau)$, where $L(\cdot)$ is the usual likelihood 
function and $\widehat{\phi}_\tau$ is the restricted maximum likelihood 
estimator of $\phi$ given $\tau$. The profile likelihood is not a true likelihood,
and some of the properties that hold for a genuine likelihood do 
not hold for its profiled version. In particular, there exist score and 
information biases, both of order $\mathcal{O}(1)$. 

The interest lies in testing the null hypothesis $\mathcal{H}_0: \tau=\tau_{0}$ against
$\mathcal{H}_{1}: \tau \not= \tau_{0}$, where $\tau_0$ is a given $q$-vector of scalars. 
The likelihood ratio statistic obtained from the profile likelihood function 
is 
$$LR = 2\left\{ \ell(\widehat\tau,\widehat\phi) - 
\ell(\tau,\widehat\phi_\tau)\right\}=2\left\{\ell_p(\widehat\tau) - \ell_p(\tau)\right\}.$$
Here, $\widehat{\tau}$ and $\widehat{\phi}$ are the maximum likelihood estimators 
of $\tau$ and $\phi$, respectively, $\ell(\cdot)$ is the log-likelihood function and 
$\ell_p(\cdot)$ is the profile log-likelihood function. Under 
the null hypothesis, $LR \leadsto \chi^2_q$, where $\leadsto$ denotes 
convergence in distribution. 

Several adjustments to the profile likelihood function have been proposed 
in the 
literature; see, e.g., Severini (2000, Chapter 9), Pace and Salvan (1997, 
Chapter 11) and the referenced therein for details. 

Barndorff--Nielsen (1983) proposed an adjusted profile likelihood function which is 
invariant under reparameterizations of the form $(\tau,\phi) \rightarrow 
(\tau,\zeta(\tau,\phi))$, where $\tau$ is the vector of parameters of 
interest, $\phi$ is the vector of nuisance parameters and $\zeta$ is a function 
of $\tau$ and $\phi$. His proposal follows from the $p^*$ formula, which is 
an approximation to the conditional density of the maximum likelihood estimator 
given an ancillary statistic. The proposed adjusted profile likelihood function is 
$$L_{BN}(\tau)=\Biggl\vert{\partial\widehat\phi_\tau \over 
\partial\widehat\phi}\Biggr\vert^{-1}
\vert j_ {\phi\phi}(\tau,\widehat\phi_\tau) \vert^{-1/2} L_p(\tau),$$
where $\partial\widehat\phi_\tau / \partial\widehat\phi$ is the matrix of 
partial derivatives of $\widehat\phi_\tau$ with respect to $\widehat\phi$, 
$j_ {\phi\phi}(\tau,\phi)=-{\partial^2\ell/\partial\phi\partial\phi^\top}$ is the 
observed information matrix for $\phi$ when $\tau$ is fixed and $L_p(\tau)$ 
is the profile likelihood function for $\tau$.

There is an alternative expression for $L_{BN}$ that does not involve 
$\vert{\partial\widehat\phi_\tau / \partial\widehat\phi}\vert$; it involves, 
nonetheless, a sample space derivative and requires an ancillary statistic $a$ 
such that $(\widehat{\tau},\widehat{\phi},a)$ is minimal sufficient. 
It can be shown that 
$${\partial\widehat\phi_\tau \over \partial\widehat\phi} = 
j_{\phi\phi}(\tau,\widehat\phi_\tau;
\widehat\tau,\widehat\phi,a)^{-1}
\ell_{\phi;\widehat\phi}(\tau,\widehat\phi_\tau;\widehat\tau,\widehat\phi,a),$$
where $$\ell_{\phi;\widehat\phi}(\tau,\widehat\phi_\tau;\widehat\tau,\widehat\phi,a)=
{\partial\over\partial\widehat\phi}\left(\partial\ell(\tau,\widehat\phi_\tau;\widehat\tau,\widehat\phi,a)\over
\partial\phi\right).$$
Here, $\ell_{\phi;\widehat\phi}(\tau,\widehat\phi_\tau;\widehat\tau,\widehat\phi,a)$  
and $j_{\phi\phi}(\tau,\widehat\phi_\tau;
\widehat\tau,\widehat\phi,a)$ are the log-likelihood function and the observed 
information for $\phi$, respectively. They depend on the data only through the minimal 
sufficient statistic. 

Some approximations to the sample space derivative of the log-likelihood 
function have been proposed. Severini (1998) obtained an approximation to 
Barndorff--Nielsen's adjusted profile likelihood function that requires 
neither a sample space derivative nor an ancillary statistic.  It is given by
$$
\bar \ell_{BN}(\tau)=\ell_p(\tau) + {1\over2}\log\vert 
j_{\phi\phi}(\widehat\tau,\widehat\phi_\tau)\vert
- \log\vert I_\phi(\tau,\widehat\phi_\tau;\widehat\tau,\widehat\phi)\vert,$$
where 
$$I_\phi(\tau,\phi;\tau_0,\phi_0)=
\E_{(\tau_0,\phi_0)}\left\{\ell_\phi(\tau,\phi)\ell_\phi(\tau_0,\phi_0)^\top\right\} \eqno{(1)}$$
is the covariance matrix of log-likelihood derivatives and 
$\ell_\phi(\tau,\phi)=\partial \ell/ \partial\phi$.
The approximation error is of order $\mathcal{O}(n^{-1/2}).$ 
The corresponding maximum likelihood estimator shall be denoted as 
$\widehat{\bar\tau}_{BN}$.

An alternative approximation, with the same approximation error, was
proposed by Severini (1999):
$$\breve \ell_{BN}(\tau)=\ell_p(\tau) + {1\over2}\log\vert 
j_{\phi\phi}(\widehat\tau,\widehat\phi_\tau)\vert
- \log\vert \breve{I}_\phi(\tau,\widehat\phi_\tau;\widehat\tau,\widehat\phi)\vert,$$
where 
$$\breve{I}_\phi(\tau,\phi;\tau_0,\phi_0)=
\sum^n_{j=1}\ell^{(j)}_\phi(\tau,\phi)
\ell^{(j)}_\phi(\tau_0,\phi_0)^\top,\eqno{(2)}$$
$\ell_{\theta}^{(j)}(\theta)=(\ell_{\tau}^{(j)}(\theta), 
\ell_{\phi}^{(j)}(\theta)$ being the score function for the $j$th observation. 
This approximation can be easily computed and is particularly useful 
in situations where one is not able to compute expected values of 
log-likelihood derivatives. The corresponding maximum likelihood 
estimator shall be denoted as 
$\widehat{\breve\tau}_{BN}$.

Cox and Reid (1987) defined an adjusted profile likelihood function, where 
an adjustment term is included into the likelihood function prior to 
maximization. It approximates the conditional density function of the 
observations given the nuisance parameter maximum likelihood estimator and 
can be written as 
$$L_{CR}(\tau)=\vert 
j_{\phi\phi}(\tau,\widehat\phi_\tau)\vert^{-1/2}L_p(\tau).$$
Taking logs we obtain
$$\ell_{CR}(\tau)=\ell(\tau,\widehat\phi_\tau) - 
{1\over2}\log\vert j_{\phi\phi}(\tau,\widehat\phi_\tau)\vert.\eqno{(3)}$$
Note that this function is the penalized counterpart of the log-likelihood
function, the penalty term taking into account information on the nuisance 
parameter. The maximizer of $\ell_{CR}(\tau)$ shall be denoted as 
$\widehat\tau_{CR}$. It is noteworthy that the score bias is of order  
$\mathcal{O}(n^{-1})$, but the information bias remains $\mathcal{O}(1)$.

The derivation of $\ell_{CR}(\tau)$ requires that $\tau$ and $\phi$ be 
orthogonal, i.e., that the elements of the score vector, 
$\partial \ell/\partial \tau$ and $\partial \ell/\partial \phi$, be 
uncorrelated which implies that $i_{\tau\phi}=0$. When $i_{\tau\phi}\not=0$, 
it is necessary to find a reparameterization 
of the form $(\tau,\lambda(\tau,\phi))$, where $\tau$ and $\lambda$ are 
orthogonal. It is noteworthy that such a reparameterization cannot always 
be found, except when the parameter of interest is scalar.
We also note that the Cox and Reid adjustment 
is not invariant under reparameterizations of the form 
$(\tau,\phi) \rightarrow (\tau,\zeta(\tau,\phi))$,
unlike Barndorff--Nielsen's adjustment. 

\section{The Birnbaum--Saunders adjusted profile like\-li\-hoods}\label{S:BSlikelihoods}

At the outset, let $\alpha$ be the parameter of interest and $\beta$ the 
nuisance parameter. Also, let $\mathbf{t}=(t_1,\dots,t_n)^\top$ denote a random sample 
of size $n$ from the Birnbaum--Saunders distribution. The log-likelihood function
is
$$\ell(\alpha,\beta)=-n\log(\alpha\beta) + 
\sum^{n}_{i=1}\log\left[\left({\beta\over t_i}\right)^{1/2} +
\left({\beta\over t_i}\right)^{3/2} \right] -
{1\over 2\alpha^2}\sum^{n}_{i=1}\left({t_i \over \beta} + {\beta \over t_i} - 2\right).$$

For fixed $\alpha$, the restricted maximum likelihood estimator of $\beta$, 
$\widehat\beta_\alpha$, is the positive root of the following nonlinear 
equation:
$$\beta^2 - \beta[2r + K(\beta)] + r[s + K(\beta)] = 0,$$
where
$$s={1\over n}\sum^n_{i=1}t_i,\quad r=\left({1\over 
n}\sum^n_{i=1}t_i^{-1}\right)^{-1}
\quad \hbox{and}\quad K(\beta)=\left[{1\over n} \sum^n_{i=1}(\beta + t_i)^{-1}\right]^{-1}.$$
Note that $\widehat\beta_\alpha$ does not have a closed-form expression,  
and, as a result, it must be obtained using restricted nonlinear optimization 
methods; 
see, e.g., Nocedal and Wright (1999, Chapter 18). (We note that the maximum likelihood 
estimator of $\beta$ for fixed $\alpha$ equals the maximum likelihood estimator of 
$\beta$, that is, $\widehat\beta_\alpha = \widehat\beta$.) 
By replacing $\beta$ by
$\widehat\beta_{\alpha}$ in $\ell(\alpha,\beta)$ we obtain the profile log-likelihood 
function given by
\begin{eqnarray*}
\ell_p(\alpha)&=&-n\log\alpha - n\log\widehat\beta_{\alpha} + \sum^{n}_{i=1}
\log\left[\left({\widehat\beta_{\alpha}\over t_i}\right)^{1/2} +
\left({\widehat\beta_{\alpha}\over t_i}\right)^{3/2} \right]\\ 
&-& {n \over 2\alpha^2}\left({r\over\widehat\beta_{\alpha}}
+{\widehat\beta_{\alpha}\over s}-2 \right).
\end{eqnarray*}

The asymptotic distribution of the vector of maximum likelihood estimators 
of the parameters that index the Birnbaum--Saunders distribution was  
obtained by Englehardt, Bain and Wright~(1981). A simpler expression for 
Fisher's information matrix was obtained by Lemonte, Cribari--Neto and 
Vasconcellos~(2007). 

In what follows, we shall obtain the adjusted profile likelihoods described in 
Section \ref{S:likelihoods}. Note that the interest and nuisance parameters are orthogonal. The 
adjusted profile log-likelihood function of Cox and Reid (1987) for $\alpha$ 
can be expressed as  
$$\ell_{CR}(\alpha)=\ell_p(\alpha) - {1\over2}\log\vert 
j_{\beta\beta}(\alpha,\widehat\beta_\alpha)\vert,$$
where
$$j_{\beta\beta}(\alpha,\widehat\beta_\alpha)=-{n\over \widehat\beta_\alpha^2}+
{n\over2}\left({1\over\widehat\beta_{\alpha}^2} + {2K'(\widehat\beta_\alpha)\over 
K^2(\widehat\beta_{\alpha})}
\right) + {n\over\alpha^2}{r\over\widehat\beta_{\alpha}^3}$$
and
$$
K'(\beta)={n\sum^n_{i=1}(\beta + t_i)^{-2} \over \left[\sum^n_{i=1}(\beta + t_i)^{-1}\right]^2}.$$
$\widehat{\,\!\alpha}_{CR}$ is the adjusted profile maximum likelihood estimator 
of $\alpha$; it does not have closed-form and must be obtained numerically. 

The Barndorff--Nielsen (1983) adjusted profile log-likelihood function 
for $\alpha$ can be written as 
$$\ell_{BN}(\alpha)=\ell_{p}(\alpha) + \log{\vert 
j_{\beta\beta}(\alpha,\widehat\beta_\alpha) \vert^{1/2}  \over \vert 
j_{\beta;\widehat\beta}(\alpha,\widehat\beta_\alpha)\vert}.$$
Instead of obtaining the term $j_{\beta;\widehat\beta}(\alpha,\widehat\beta_\alpha)$ 
in $\ell_{BN}$, we shall obtain 
$\breve I(\alpha,\,\widehat{\!\beta}_\alpha;\widehat{\alpha},\,\widehat{\!\beta})$ 
given in (2) using, thus, Severini's (1999) approximation. After some algebra, 
we obtain
\begin{eqnarray}
\breve I(\alpha,\,\widehat{\!\beta}_\alpha;\widehat{\alpha},\,\widehat{\!\beta})&=&
{n\over\widehat\beta^2}-{1\over\widehat\beta}\sum^n_{j=1}A_{j}
-{1\over 4}\left[ \sum^n_{j=1}A_{j}^2 + {1\over\alpha^2\widehat\alpha^2}\sum^n_{j=1}B_{j}^2 \right] \nonumber \\
&-&{1\over 4}\left[\left({1\over\alpha^2}+{1\over\widehat\alpha^2}\right)
\left(\sum^n_{j=1}A_{j}B_{j}-{2\over\widehat\beta}\sum^n_{j=1}B_{j}\right)
\right], \nonumber
\end{eqnarray}
where
$$A_j=\left({t_j^{-1/2}\widehat\beta^{-1/2}+3\widehat\beta^{1/2}t_j^{-3/2}\over
t_j^{-1/2}\widehat\beta^{1/2}+\widehat\beta^{3/2}t_j^{-3/2}}\right) 
\qquad\hbox{and}\qquad
B_j=\left({t_j\over \widehat\beta^2}-{1\over t_j}\right).$$
The adjusted profile maximum likelihood estimator $\widehat{\,\!\alpha}_{BN}$ of
$\alpha$ cannot be expressed in closed-form; it has to computed by numerically 
maximizing the associated log-likelihood function. 

The likelihood ratio test statistics obtained from the adjusted profile 
log-likelihood functions given above for the test of 
$\mathcal{H}_0: \alpha=\alpha_0$ against $\mathcal{H}_1: \alpha\not=\alpha_0$ are 
$${LR}_{CR}(\alpha) = 2\left\{\ell_{CR}(\widehat\alpha_{CR}) - \ell_{CR}(\alpha_0)\right\}$$
and
$${LR}_{BN}(\alpha) = 2\left\{\ell_{BN}(\widehat\alpha_{BN}) - \ell_{BN}(\alpha_0)\right\},$$
where $\widehat\alpha_{CR}$ and $\widehat\alpha_{BN}$ are the values of $\alpha$ that maximize 
$\ell_{CR}(\alpha)$ and $\ell_{BN}(\alpha)$, respectively.
These test statistics are asymptotically distributed as $\chi_1^2$ under 
the null hypothesis. 

We shall now consider $\beta$ as the parameter of interest and view $\alpha$ as
a  
nuisance parameter. For fixed $\beta$, we write the restricted maximum likelihood 
estimator of $\alpha$ as 
$$\widehat\alpha_\beta=\left({r \over\beta} + {\beta \over s} - 
2\right)^{1/2}.$$
By plugging $\widehat\alpha_\beta$ into the log-likelihood function we obtain 
the following profile log-likelihood function:
\begin{eqnarray*}
\ell_p(\beta)&=&\ell(\widehat\alpha_\beta,\beta)=-{n\over2}\log\left({r\over\beta}+{\beta\over 
s}-2\right)-
n\log\beta\\ &+& \sum_{i=1}^n\log\left[\left({\beta\over t_i}\right)^{1/2}
+ \left({\beta\over t_i}\right)^{3/2} \right].
\end{eqnarray*}

The $j_{\alpha\alpha}(\alpha,\beta)$ block of the observed information matrix
evaluated at $(\widehat\alpha_\beta,\beta)$ can be written as 
$$j_{\alpha\alpha}(\widehat\alpha_\beta,\beta)=-{2n \left({r\over\beta} + 
{\beta\over s} - 2\right)^{-1}}.$$
From (3) it follows that Cox and Reid's adjusted profile log-likelihood 
function for $\beta$ is 
$$\ell_{CR}(\beta)=\ell_p(\beta) + {1\over2}\log\left\vert 
{r\over\beta}+{\beta\over s}-2 \right\vert.$$
The estimator $\widehat\beta_{CR}$, like the previous estimators, does not 
have closed-form. 

The Barndorff--Nielsen adjusted profile log-likelihood function can be expressed as
$$\ell_{BN}(\beta)=\ell_{p}(\beta) + \log {\vert 
j_{\alpha\alpha}(\widehat\alpha_\beta,\beta)\vert^{1/2}  \over \vert 
\ell_{\alpha;\widehat\alpha}(\widehat\alpha_\beta,\beta)\vert}.$$
We use Severini's (1998) approximation and replace 
$\ell_{\alpha;\widehat\alpha}(\widehat\alpha_\beta,\beta)$, in $\ell_{BN}(\beta)$, 
by $I(\widehat\alpha_\beta,\beta;\widehat{\,\!\alpha},\widehat{\,\!\beta})$ given in 
(1). We arrive at 
$$I(\widehat\alpha_\beta,\beta;\widehat{\,\!\alpha},\widehat{\,\!\beta})=
{n\widehat{\,\!\alpha}\over\widehat{\,\!\alpha}^3_\beta}
\left({\,\widehat{\!\beta}\over\beta} + {\beta\over\,\widehat{\!\beta}} 
\right).$$ The corresponding estimator, $\,\widehat{\!\beta}_{BN}$, does not have 
closed-form. 

The likelihood ratio test statistics obtained from the above adjusted profile 
log-likelihood functions for the test of $\mathcal{H}_0: \beta=\beta_0$ against
$\mathcal{H}_1: \beta\not=\beta_0$ are 
$${LR}_{CR}(\beta) = 2\left\{\ell_{CR}(\widehat\beta_{CR}) - 
\ell_{CR}(\beta_0)\right\}$$
and 
$${LR}_{BN}(\beta)=2\left\{\ell_{BN}(\widehat\beta_{BN}) - \ell_{BN}(\beta_0)\right\},$$
where $\widehat\beta_{CR}$ and $\widehat\beta_{BN}$ are the values of $\beta$ that maximize $\ell_{CR}(\beta)$ and
$\ell_{BN}(\beta)$, respectively. The two test statistics are asymptotically 
distributed as $\chi_1^2$ under $\mathcal{H}_0$. 

\section{Alternative inference strategies}\label{S:alternativeestimators}

Some alternative point estimators for the parameters that index the 
Birnbaum--Saunders distributions have been proposed in the literature. 
Ng, Kundu and Balakrishnan~(2003) obtained modified moment estimators 
for $\alpha$ and $\beta$. As before, let $s = \overline{t} = n^{-1} 
\sum_{i=1}^n t_i$ (sample arithmetic mean) and 
$r = \left( n^{-1} \sum_{i=1}^n t_i^{-1} \right)^{-1}$ (sample harmonic 
mean). The estimators 
can then be written as   
\begin{equation*}
\bar\alpha_{Ng} = \sqrt{s \left( \sqrt{\frac{s}{r} - 1}\right) } 
\quad \mathrm{and} \quad \bar\beta_{Ng} = \sqrt{sr}. 
\end{equation*}

Ng, Kundu and Balakrishnan (2003) have also proposed jackknife estimators for 
$\alpha$ and $\beta$. The underlying idea is to remove observation $t_j$ from 
the random sample ${\bf t} = (t_1, t_2, \ldots, t_n)^\top$, and to estimate 
the parameters based on the remaining $n-1$ observations; this is done for 
$j=1,\ldots,n$. We shall denote the jackknife maximum likelihood estimators 
as $\bar\alpha_{Ng_{JMLE}}$ and $\bar\beta_{Ng_{JMLE}}$; the jackkinife 
moment estimator are $\bar\alpha_{Ng_{JMME}}$ and $\bar\beta_{Ng_{JMME}}$. 

From and Li~(2006) also proposed alternative estimators for the two 
parameters that index the Birnbaum--Saunders distribution. For instance, 
they proposed using 
\begin{equation*}
\breve\beta_{F1} = \frac{\sum_{i=1}^n t_i^{1/2}}{\sum_{i=1}^n t_i^{-1/2}}
\quad \mathrm{and} \quad  
\breve\alpha_{F1} = \sqrt{ \frac{s}{\breve\beta_{F1}}
+ \frac{\breve\beta_{F1}}{r} - 2 }.  
\end{equation*}
The authors have also proposed a second estimator for $(\alpha,\beta)$. 
Their proposal is to estimate $\beta$ using $\breve\beta_{F2} = 
\mathrm{median}(t_1,\ldots,t_n)$ since $\beta$ equals the median of the 
Birnbaum--Saunders distribution. The corresponding estimator for $\alpha$ is 
\begin{equation*}
\breve\alpha_{F2} = \sqrt{\frac{-2+2\sqrt{1+5v}}{5}}, 
\end{equation*}
where $v = {\widehat{\sigma}^2}/{\breve\beta_{F2}}$, 
$\widehat{\sigma}^2$ being the sample variance,
i.e., $\widehat{\sigma}^2 = (n-1)^{-1}\sum_{i=1}^n (t_i - \overline{t})^2$.

Let $t_{(1)},\ldots,t_{(n)}$ denote the order statistics of the sample $t_1,\ldots,
t_n$. The estimator for $\beta$ is, as above, the sample median.
For each $t_{(i)}$,
solve 
\begin{equation*}
F(t_{(i)}; \alpha, \breve\beta_{F2}) = \frac{i}{n+1}, \quad i=1,\ldots, n. 
\end{equation*}
Let $\widehat{\alpha}(i)$, $i=1,\ldots,n$, denote the solutions, where 
\begin{equation*}
\widehat{\alpha}(i) = \frac{h\left(\frac{t_{(i)}}{\breve\beta_{F2}}\right)}{\Phi^{-1}\left(\frac{i}{n+1}\right)},
\end{equation*}
with $h(t) = t^{1/2} - t^{-1/2}$. The estimator is 
$\breve\alpha_{F3} = \mathrm{median}(\widehat{\alpha}(1), \ldots, \widehat{\alpha}(n))$.

From and Li~(2006) have also proposed yet another estimator for $(\alpha,\beta)$. 
Let $0 < \lambda < 0.5$, and let $n_1 = n \lambda + 1$ and $n_2 = n(1-\lambda)$,
to the nearest integer. The proposed estimators are 
\begin{equation*}
\breve\beta_{F4} =  \frac{\sum_{i=n_1}^{n_2} t_{(i)}}{\sum_{i=n_1}^{n_2}
\frac{1}{\sqrt{t_{(i)}}}}
\quad \mathrm{and} \quad 
\breve\alpha_{F4} = \sqrt{\frac{\sum_{i=n_1}^{n_2}h^2\left( \frac{t_{(i)}}{\breve\beta_{F2}}\right)} 
{\sum_{i=n_1}^{n_2}\left[ \Phi^{-1}\left( \frac{i}{n+1} \right)\right]}}.
\end{equation*}
The authors suggest using $\lambda = 0.05$, so that only the middle 90\% of 
order statistics are used. 

An alternative hypothesis test was proposed by Lemonte, Cribari--Neto and 
Vasconcellos~(2007). They derived a Bartlett-correction factor to the 
likelihood ratio statistic and obtained an approximate test whose error 
vanishes at a faster rate as the sample size increases. Let $LR^*$ denote 
their test statistic. It follows that whereas $\Pr(LR \leqslant x) = 
\Pr(\chi_{1}^{2} \leqslant x) + \mathcal{O}(n^{-1})$, the correction yields 
$\Pr(LR^* \leqslant x) = \Pr(\chi_{1}^{2} \leqslant x) + \mathcal{O}(n^{-2})$, 
a clear improvement. 
[See Cribari--Neto and Cordeiro~(1996) for a detailed review of Bartlett 
corrections.] Consider the following null hypotheses: 
(i) ${\mathcal H_{0}}\!: \alpha = 0.1$, 
(ii) ${\mathcal H_{0}}\!: \alpha = 0.25$,
(iii) ${\mathcal H_{0}}\!: \alpha = 0.5$,
(iv) ${\mathcal H_{0}}\!: \alpha = 0.75$,
(v) ${\mathcal H_{0}}\!: \alpha = 1.0$, 
(v) ${\mathcal H_{0}}\!: \alpha = 2.0$. 
The corresponding Bartlett-corrected test statistics are\footnote{Note that 
the values of the Bartlett correction factor we give correct those given 
by the authors for cases (i) and (ii).}  
\[
LR^* = \frac{LR}{1 + 4.3918/n},\ LR^* = \frac{LR}{1 + 3.2537/n},\ LR^* = \frac{LR}{1 + 3.0414/n},
\]
\[
LR^* = \frac{LR}{1 + 2.5924/n},\ LR^* = \frac{LR}{1 + 2.0307/n}
\ \ \mathrm{and} \ \ LR^* = \frac{LR}{1 - 0.0445/n}.
\]

\section{Numerical evidence}\label{S:MonteCarlo1}

We shall now present Monte Carlo simulation results on the finite sample 
behavior of inference based on profile and adjusted profile likelihoods. 
All simulation experiments entail 10,000 replications.
The shape parameter assumed two values, namely 
$\alpha=0.5,1.0$, and the scale parameter was set at $\beta=1.0$. 
The simulations were performed using the \texttt{Ox} matrix programming 
language 
(Doornik, 2006). Likelihood maximizations were performed using 
the quasi-Newton method known as BFGS with analytical first derivatives;
see Nocedal and Wright~(1999) for details on the BFGS method.  

Point estimation is evaluated using the following measures: mean, bias, 
variance, mean squared error (MSE), relative bias (RB), asymmetry and 
kurtosis. Relative bias is defined as  
$100 \times (\mbox{bias} / \mbox{true parameter value})$.
Hypothesis testing inference on the parameter of interest is described
through the null rejection rates of the profile and adjusted profile 
likelihood ratio tests. Power simulations were also performed. 

Table 1 contains simulation results for the case where $\alpha$ is the 
parameter of interest. The sample size is $n=10$. Note that the 
estimators $\widehat\alpha_{CR}$ 
and $\widehat\alpha_{BN}$ are less biased than $\widehat\alpha_{p}$. 
For instance, when $\alpha=0.5$ the relative biases of $\widehat\alpha_{p}$, 
$\widehat\alpha_{CR}$ and $\widehat\alpha_{BN}$ are 7.50\%, 2.16\%  and 2.13\%,
respectively.
Nevertheless, bias reduction is achieved at the expense of greater 
variability. 
It is also noteworthy that the small sample behavior of the two adjusted profile 
maximum likelihood estimators are similar. 
%When $\alpha=1$, $\widehat\alpha_{p}$ 
%is clearly outperformed by the adjusted profile maximum likelihood estimators, 
%the best performing estimator being $\widehat\alpha_{BN}$. 
We also note 
that the skewness and kurtosis of $\widehat\alpha_{p}$ are slightly closer to their 
asymptotic counterparts than those of $\widehat\alpha_{CR}$ and $\widehat\alpha_{BN}$.
(When the parameter of interest is $\beta$, the estimators $\widehat\beta_p$, 
$\widehat\beta_{CR}$ and $\widehat\beta_{BN}$ coincide, since maximization of 
of the profile likelihood function is equivalent to that of 
$\ell_{CR}(\beta)$ or $\ell_{BN}(\beta)$. As a consequence, the profile and 
adjusted profile maximum likelihood estimators also coincide.) 

%ao estimador $\widehat\beta$. Observamos que este estimador apresentou 
%comportamento bastante
%satisfatório, entretanto, para valores elevados do parâmetro $\alpha$ 
%observamos uma deterioração
%no seu desempenho.

\begin{sidewaystable}
\begin{center}
\caption{Point estimation of $\alpha$.}
\vspace{0.13cm}
\begin{tabular}{c c c c c c c c}\hline
\multicolumn{8}{c}{$\alpha=0.5$}\\ \hline
estimator           &mean &bias    &variance & MSE      &RB(\%)    
&asymmetry &kurtosis\\ \hline
$\widehat\alpha_p$   &0.4625 & $-0.0375$ &   0.0121 &   0.0135&   7.5074&   
0.2194&   6.0405\\
$\widehat\alpha_{CR}$&0.4892 & $-0.0109$ &   0.0136 &   0.0138&   2.1695&   
0.2355&   6.3629\\
$\widehat\alpha_{BN}$&0.4893 & $-0.0107$ &   0.0137 &   0.0138&   2.1385&   
0.2356&   6.3648 \\ 
\end{tabular}

\begin{tabular}{c c c c c c c c}\hline
\multicolumn{8}{c}{$\alpha=1.0$}\\ \hline
estimator           &mean &bias    &variance & MSE      &RB(\%)    
&asymmetry &kurtosis\\ \hline
$\widehat\alpha_p$   &0.9160&  $-0.0840$&   0.0472&   0.0543&   8.4020&   
0.3715&  11.9221\\
$\widehat\alpha_{CR}$&0.9752&  $-0.0248$&   0.0548&   0.0554&   2.4841&   
0.3732&  12.6356\\
$\widehat\alpha_{BN}$&0.9792&  $-0.0208$&   0.0567&   0.0572&   2.0834&   
0.3733&  12.6829\\ \hline
\end{tabular}
\end{center}
\end{sidewaystable}

Table 2 presents the null rejection rates (\%)
of the different likelihood ratio tests, 
i.e., the tests based on the statistics $LR$, $LR_{CR}$, $LR_{BN}$ and $LR^*$, 
for 
the test of $\mathcal{H}_0:\alpha=\alpha_0$ against $\mathcal{H}_1:\alpha\not=\alpha_0$, where 
$\alpha_0$ is a given scalar; here, $\alpha$ is the parameter of interest and 
$\beta$ is the nuisance parameter, whose value is set at $\beta=1.0$. The 
values set at the null hypothesis are $\alpha_0 = 0.5, 1.0, 2.0$ and the sample 
sizes are $n=10, 25, 50$. All entries are percentages. The figures in Table 2 
show that the adjusted profile likelihood ratio tests ($LR_{CR}$ and $LR_{BN}$) 
outperform the usual profile likelihood ratio test ($LR$). For instance, at the 
10\% nominal level and when $\alpha_0 = 0.5$, the rejection rate of the latter was 
13.57\% whereas the rejection rates of $LR_{CR}$ and $LR_{BN}$ were 10.85\%
and 10.86\%, respectively. The likelihood ratio test is clearly liberal, the 
null hypothesis being rejected more often than expected. The adjusted tests 
display much smaller size distortions than the likelihood ratio test.  
It is also noteworthy 
that the finite sample behavior of the likelihood ratio test deteriorates as 
the value of the shape parameter increases, especially when the sample size 
is small; the adjusted tests remain reliable. 
The Bartlett-corrected test is clearly outperformed by the adjusted profile 
likelihood tests when $\alpha$ is large and $n$ is small. For example, 
when $\alpha=2.0$, $n=10$ and the nominal level is $5\%$, the null rejection 
rate of the Bartlett-corrected test is 9.40\% whereas the adjusted profile 
likelihood tests reject the null hypothesis 5.20\% ($LR_{CR}$) and 4.58\% 
($LR_{BN}$) of the time. Note also that the tests based on 
$LR_{CR}$ and $LR_{BN}$ display similar small sample behavior, especially when 
the value of the shape parameter is small. (Values of $\alpha$ between 0.1 and 
0.5 are common in fatigue studies.)

{\footnotesize
\begin{sidewaystable}
\begin{center}
\caption{Null rejection rates, inference on $\alpha$ ($\beta = 1.0$).}
\vspace{0.03cm}
\begin{tabular}{c | c | c c c c | c c c c | c c c c | c c c c}\hline
     & 
nominal&\multicolumn{4}{c|}{$\alpha=0.1$}&\multicolumn{4}{c|}{$\alpha=0.5$}&\multicolumn{4}{c|}{$\alpha=1.0$}&\multicolumn{4}{c}{$\alpha=2.0$}\\ 
\cline{3-18}
$n$  &{level}& 
$LR$&$LR_{CR}$&$LR_{BN}$&$LR^*$&
$LR$&$LR_{CR}$&$LR_{BN}$&$LR^*$&
$LR$&$LR_{CR}$&$LR_{BN}$&$LR^*$&
$LR$&$LR_{CR}$&$LR_{BN}$&$LR^*$\\ 
\hline
     & 10 & 13.44 & 11.00 & 11.00 & 7.24 & 13.57 & 10.85 & 10.86 & 9.03 & 13.97 & 10.65 & 10.78 & 10.61 & 16.10 &  9.89 &8.99& 15.90\\
10   & 5  &  7.47 &  5.14 &  5.14 & 3.05 &  7.60 &  5.18 &  5.20 & 3.96 &  7.83 &  5.24 &  5.29 &  5.11 &  9.44 &  5.20 &  4.58 &  9.40 \\
     & 1  &  1.70 &  1.07 &  1.07 & 0.45 &  1.72 &  1.11 &  1.11 & 0.69 &  1.83 &  1.06 &  1.11 &  0.93 &  2.53 &  1.04 &  1.96 &  2.51\\
     & 0.5&  0.89 &  0.53 &  0.53 & 0.16 &  0.91 &  0.52 &  0.52 & 0.31 &  0.97 &  0.52 &  0.55 &  0.51 &  1.40 &  0.48 &  0.40 &  1.39\\ \hline
     & 10 & 10.84 & 10.15 & 10.15 & 8.42 & 10.88 & 10.13 & 10.14 & 9.15 & 11.56 & 10.41 & 10.50 & 10.30 & 11.98 & 10.12 & 10.10 &  11.96\\
25   & 5  &  5.94 &  5.33 &  5.33 & 4.32 &  5.93 &  5.34 &  5.35 & 4.67 &  6.18 &  5.31 &  5.33 &  5.28 &  6.24 &  5.36 &  5.30 &  6.22\\
     & 1  &  1.43 &  1.18 &  1.18 & 0.83 &  1.46 &  1.15 &  1.15 & 1.02 &  1.46 &  1.04 &  1.04 &  1.04 &  1.54 &  1.07 &  1.06 &  1.54\\
     & 0.5&  0.82 &  0.57 &  0.57 & 0.43 &  0.82 &  0.59 &  0.59 & 0.48 &  0.61 &  0.50 &  0.52 &  0.43 &  0.82 &  0.59 &  0.54 &  0.81 \\ \hline
     & 10 & 10.89 & 10.35 & 10.35 & 9.46 & 10.86 & 10.35 & 10.34 & 9.99 & 10.93 & 10.31 & 10.32 & 10.25 & 11.24 & 10.48 & 10.47 & 11.23\\
50   & 5  &  5.54 &  5.21 &  5.21 & 4.58 &  5.53 &  5.20 &  5.19 & 4.81 &  5.40 &  5.03 &  5.05 &  4.86 &  5.65 &  4.95 &  5.04 &  5.64\\
     & 1  &  1.19 &  0.99 &  0.99 & 0.87 &  1.21 &  1.02 &  1.02 & 0.99 &  1.14 &  1.02 &  1.02 &  1.10 &  1.18 &  1.03 &  1.03 &  1.17\\
     & 0.5&  0.70 &  0.56 &  0.56 & 0.51 &  0.71 &  0.55 &  0.55 & 0.58 &  0.68 &  0.61 &  0.61 &  0.61 &  0.60 &  0.49 &  0.51 &  0.60\\ \hline 
\end{tabular}                           
\end{center}                            
\end{sidewaystable}                     
}

We have also performed simulations under the alternative hypothesis. The powers 
of the tests $LR$, $LR_{CR}$, $LR_{BN}$ and $LR^*$ at the 5\% 
and 1\% nominal levels 
were computed for values of $\alpha$ ranging from 0.12 to 0.28, the null 
hypothesis under test being $\mathcal{H}_0: \alpha = \alpha_0$. The tests were carried out 
using exact critical values, which were estimated in the size simulations. 
This was done so that the different tests have the same size, and  
power comparisons become meaningful. The results are presented in Table 3
and were obtained using $n=10$, $\alpha=0.10$ and $\beta=1.0$. 
(All entries are percentages.)
We note that $LR$ is slightly less powerful than $LR_{BN}$
and $LR_{CR}$. For example, when $\alpha=0.20$ and at the 5\% 
nominal level, the nonnull rejection rates of these tests were equal to 
77.86\%, 83.81\% and 83.81\%, respectively; the nonnull rejection rate 
of the Bartlett-corrected test was 77.87\%.  

%\begin{table}
%\begin{center}
%\caption{Nonnull rejection rates, inference on $\alpha$.}
%\vspace{0.13cm}
%\begin{tabular}{c | c c c | c c c}\hline
%             & \multicolumn{3}{c|}{5\%}   &  \multicolumn{3}{c}{1\%}  \\ 
%\cline{2-7}
%$\alpha$           & $ LR $&$ LR_{CR}$&$ LR_{BN}$&$ LR  $&$ LR_{CR}$&$ 
%LR_{BN}$\\ \hline
% 0.12 &    8.33 &  12.82 &  12.72 &   2.53 &   4.42 &   4.40 \\       
% 0.14 &   24.26 &  33.25 &  33.18 &  11.96 &  18.00 &  17.92 \\       
% 0.16 &   45.28 &  55.35 &  55.28 &  28.68 &  37.81 &  37.72 \\       
% 0.18 &   62.45 &  70.99 &  70.83 &  46.38 &  55.34 &  55.26 \\       
% 0.20 &   76.65 &  82.83 &  82.76 &  63.91 &  71.50 &  71.43 \\       
% 0.22 &   85.54 &  89.78 &  89.72 &  76.08 &  81.84 &  81.79 \\       
% 0.24 &   91.76 &  94.16 &  94.14 &  85.28 &  89.49 &  89.44 \\       
% 0.26 &   94.62 &  96.24 &  96.24 &  89.99 &  92.81 &  92.80 \\       
% 0.28 &   96.60 &  97.78 &  97.77 &  93.47 &  95.46 &  95.45 \\ \hline
%\end{tabular}
%\end{center}
%\end{table}

\begin{table}
\begin{center}
\caption{Nonnull rejection rates, inference on $\alpha$.}
\vspace{0.13cm}
\begin{tabular}{c | c c c c | c c c c}\hline
             & \multicolumn{4}{c|}{nominal level: 5\%}   &  \multicolumn{4}{c}{nominal level: 1\%}  \\ 
\cline{2-9}
$\alpha$     & $ LR $&$ LR_{CR}$&$ LR_{BN}$&$LR^*$&$ LR  $& $LR_{CR}$&$ LR_{BN}$&$LR^*$ \\ \hline
 0.12 &    8.55 &  13.13 &  13.13 &  8.56  &   2.50 &   4.87 &   4.87 & 2.50 \\       
 0.14 &   24.79 &  34.07 &  34.07 & 24.79  &  11.71 &  18.01 &  18.01 & 11.71 \\       
 0.16 &   45.74 &  56.06 &  56.06 & 45.75  &  29.37 &  38.16 &  38.16 & 29.37  \\       
 0.18 &   64.78 &  72.63 &  72.63 & 64.58  &  48.57 &  59.91 &  59.91 & 48.57  \\       
 0.20 &   77.86 &  83.81 &  83.81 & 77.87  &  65.33 &  72.58 &  72.58 & 65.33  \\       
 0.22 &   86.41 &  89.85 &  89.85 & 86.41  &  77.29 &  82.97 &  82.97 & 77.29   \\       
 0.24 &   91.25 &  94.18 &  94.18 & 91.25  &  85.55 &  88.88 &  88.88 & 85.55  \\       
 0.26 &   94.91 &  96.73 &  96.73 & 94.91  &  90.17 &  93.01 &  93.01 & 90.17   \\       
 0.28 &   96.98 &  97.95 &  97.95 & 96.98  &  93.90 &  95.40 &  95.40 & 93.90   \\ \hline
\end{tabular}
\end{center}
\end{table}

Figure 1 plots the relative quantile discrepancies of the three test 
statistics against the corresponding asymptotic quantiles. Relative 
quantile discrepancy is defined as the difference between exact (estimated 
by simulation) and asymptotic quantiles divided by the latter.  
The closer to zero the relative quantile discrepancies, the better the
approximation of the exact null distribution of the test statistic by the
limiting $\chi^2_1$ distribution. It is noteworthy that the relative 
quantile discrepancies of the adjusted test statistics are considerably 
closer to zero than those of the likelihood ratio test statistic, which 
oscillate around 18\%. The relative quantile discrepancies of the two 
adjusted test statistics are very similar. 

Figure 2 plots the relative size distortions against the corresponding 
nominal levels of the tests. Relative size distortion is defined as 
the difference between $p$-values (estimated by simulation) and 
nominal levels divided by the latter. Note that the relative size 
distortion of the likelihood ratio test increases rapidly as the 
nominal level of the test decreases, which does not occur for the 
adjusted tests. Note also that the relative size distortions of the 
two adjusted tests are very similar. 

\begin{figure}[hp]
\label{fig1}
\centering
\includegraphics[width=110mm,height=75mm]{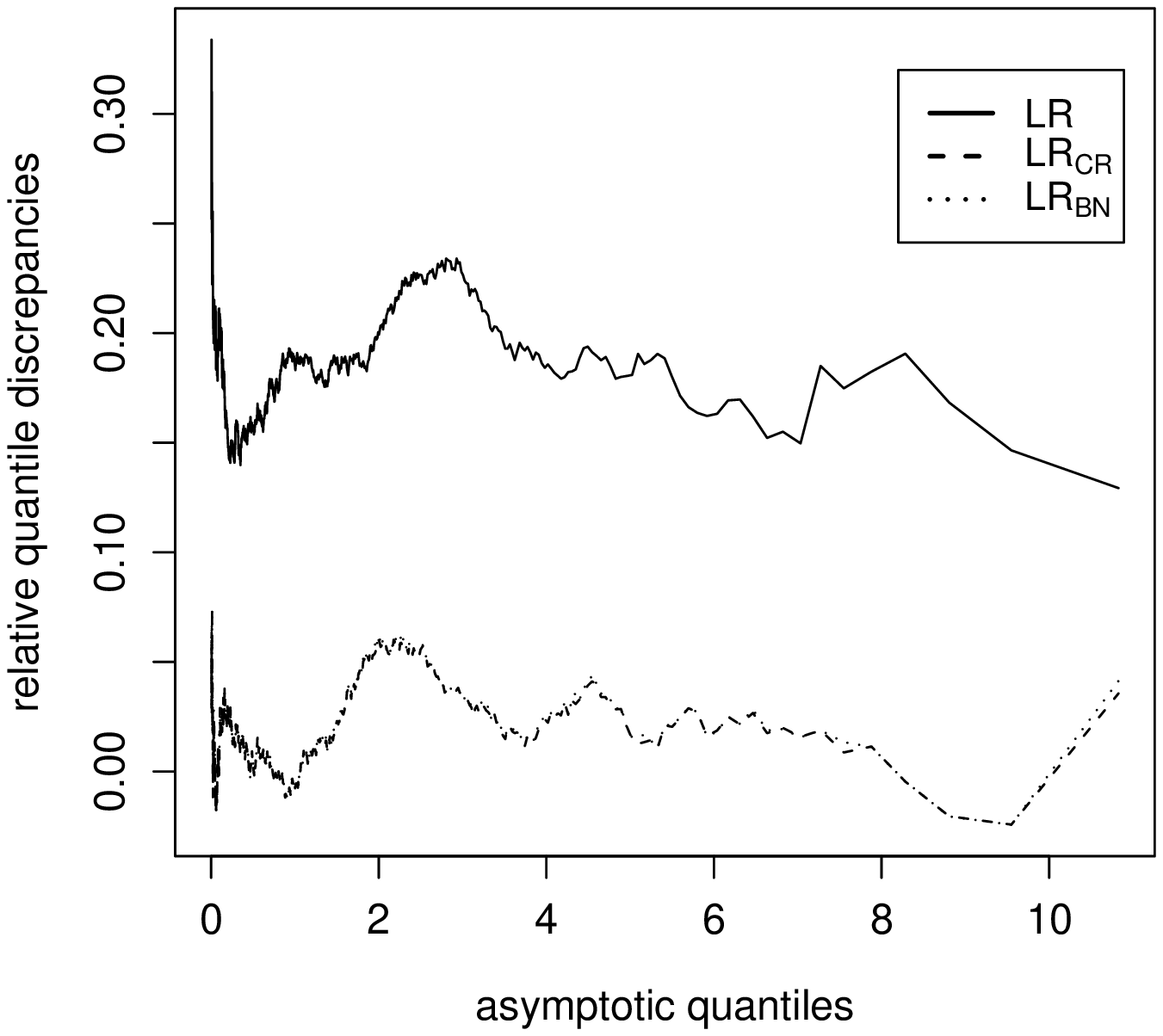}
\caption{Relative quantile discrepancy plot, inference on $\alpha$.}
\includegraphics[width=110mm,height=75mm]{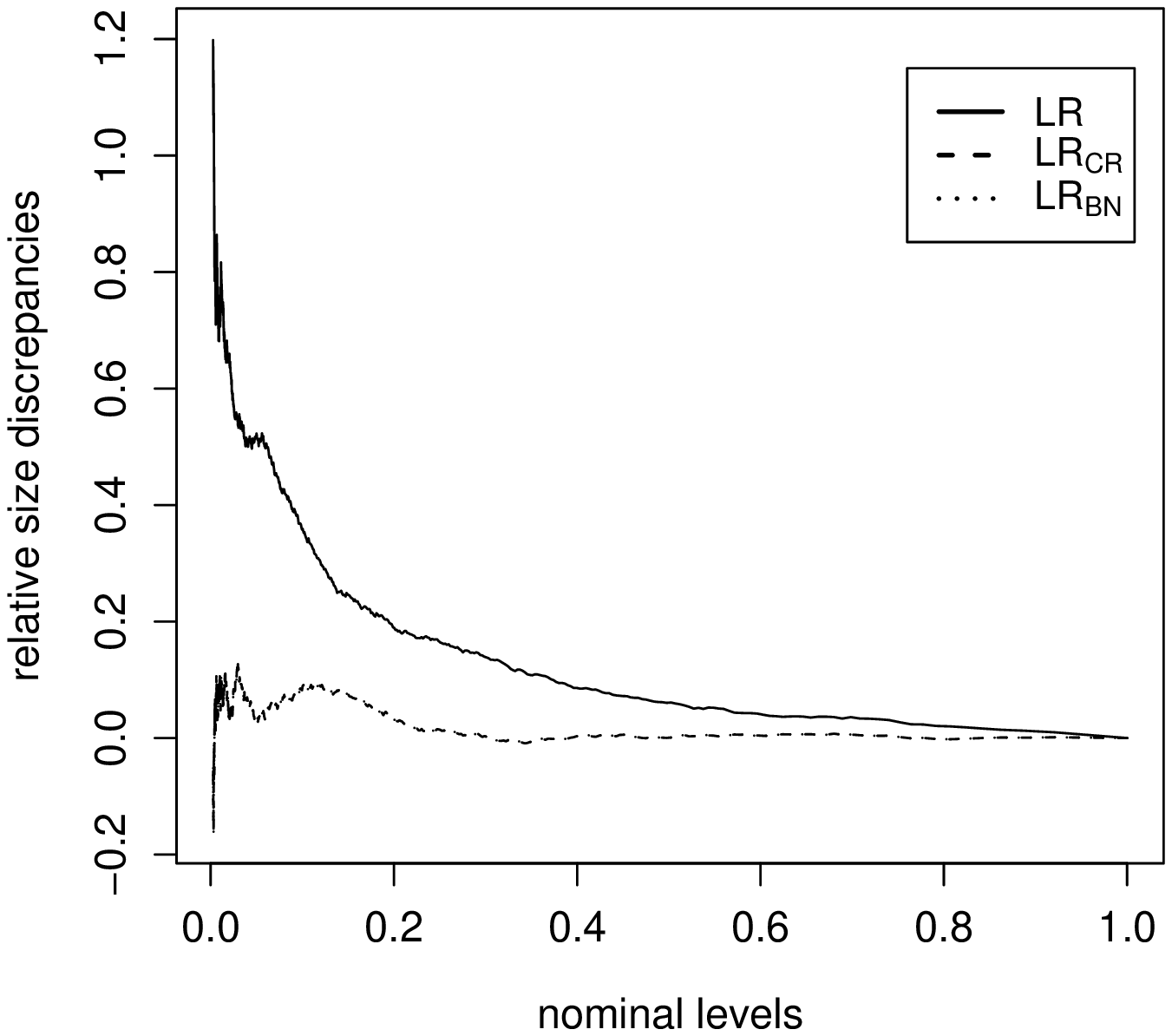}
\caption{Relative size distortion plot, inference on $\alpha$.}
\end{figure}

Table 4 contains the null rejection rates (again expressed as percentages)
of the three tests ($LR$, $LR_{CR}$ and $LR_{BN}$) for the test 
$\mathcal{H}_0:\beta=\beta_0$. (Note that here $\beta$ is the parameter of interest.) 
Since $\beta$ functions as a multiplier, as explained earlier, we have 
only performed simulations using $\beta=1.0$; four different values of 
$\alpha$ were used, namely: 0.1, 0.5, 1.0 and 2.0. As in Table 2, three 
different sample sizes were considered: $n=10, 25, 50$. The figures in Table 
4 reveal that the likelihood ratio test is liberal, that $LR_{BN}$ is 
conservative, and that $LR_{CR}$ displays very minor size distortions, the 
latter clearly outperforming the other tests. For instance, when 
$n=10$, $\alpha=2.0$ and at the 10\% nominal level, the null rejection 
rates of these tests are, respectively, 13.12\%, 7.94\% and 10.66\%.   
%As expected, the tests behave more similarly as the sample size 
%increases. 

\begin{sidewaystable}
\begin{center}
\caption{Null rejection rates, inference on $\beta$ ($\beta = 1.0$).}
\vspace{0.03cm}
{\begin{tabular}{c | c | c c c | c c c | c c c | c c c}\hline
     & 
nominal&\multicolumn{3}{c|}{$\alpha=0.1$}&\multicolumn{3}{c|}{$\alpha=0.5$}&\multicolumn{3}{c|}{$\alpha=1.0$}&\multicolumn{3}{c}{$\alpha=2.0$}\\ 
\cline{3-14}
$n$  &{level}& $LR$ 
&$LR_{CR}$&$LR_{BN}$&$LR$&$LR_{CR}$&$LR_{BN}$&$LR$&$LR_{CR}$&$LR_{BN}$&$LR$&$LR_{CR}$&$LR_{BN}$\\ 
\hline
     & 10 & 12.31 & 10.30 & 8.52 & 12.33 & 10.25 & 8.26 & 12.37 & 10.18 & 
8.01 & 13.12& 10.66&7.94\\
10   & 5  &  6.61 &  5.35 & 3.95 &  6.62 &  5.30 & 3.85 &  6.75 &  5.18 & 
3.65 &  6.95&  5.27&3.65\\
     & 1  &  1.49 &  1.08 & 0.70 &  1.56 &  1.05 & 0.69 &  1.55 &  1.09 & 
0.67 &  1.73&  1.12&0.54\\
     & 0.5&  0.91 &  0.59 & 0.35 &  0.89 &  0.60 & 0.33 &  0.87 &  0.59 & 
0.28 &  0.97&  0.53&0.25\\ \hline
     & 10 & 11.10 & 10.38 & 9.67 & 11.19 & 10.33 & 9.60 & 11.15 & 10.37 & 
9.52 & 11.27& 10.52&9.53\\
25   & 5  &  5.83 &  5.30 & 4.86 &  5.91 &  5.38 & 4.76 &  5.87 &  5.33 & 
4.75 &  6.14&  5.49&4.77\\
     & 1  &  1.25 &  1.08 & 0.93 &  1.30 &  1.07 & 0.99 &  1.40 &  1.15 & 
0.97 &  1.19&  1.09&0.96\\
     & 0.5&  0.62 &  0.52 & 0.44 &  0.64 &  0.52 & 0.45 &  0.76 &  0.61 & 
0.46 &  0.80&  0.64&0.43\\ \hline
     & 10 & 10.71 & 10.43 & 9.96 & 10.64 & 10.35 & 9.92 & 10.72 & 10.24 & 
9.74 & 10.87& 10.38&9.87\\
50   & 5  &  5.32 &  5.05 & 4.85 &  5.36 &  5.17 & 4.88 &  5.54 &  5.17 & 
4.90 &  5.39&  5.07&4.79\\
     & 1  &  1.19 &  1.08 & 1.00 &  1.20 &  1.10 & 0.99 &  1.18 &  1.09 & 
0.98 &  1.12&  1.06&0.97\\
     & 0.5&  0.56 &  0.51 & 0.48 &  0.58 &  0.53 & 0.46 &  0.61 &  0.55 & 
0.50 &  0.65&  0.59&0.52\\ \hline
\end{tabular} }
\end{center}
\end{sidewaystable} 

In Table 5 we present the empirical powers of the tests of  
$\mathcal{H}_0:\beta = \beta_0$. The values of $\beta$ used ranged from 1.2 to 4.0. 
Again, the tests were performed using size-corrected critical 
values (obtained from the size simulations) in order to force them  
to have the correct size. The simulations were carried out using 
$n=10$, $\alpha=1.0$ and $\beta=1.0$. The results suggest that the 
powers of the three tests are very similar, with a slight advantage of
$LR$. For example, when $\beta = 2.0$, the nonnull rejection rates 
of $LR$, $LR_{CR}$ and $LR_{BN}$ at the 5\% nominal level are, respectively, 
58.54\%, 58.05\% and 57.43\%.

\begin{table}
\begin{center}
\caption{Nonnull rejection rates, inference on $\beta$.}
\vspace{0.13cm}
\begin{tabular}{c | c c c | c c c}\hline
             & \multicolumn{3}{c|}{nominal level: 5\%}   &  \multicolumn{3}{c}{nominal level: 1\%}  \\ 
\cline{2-7}
$\beta$           & $ LR $&$ LR_{CR}$&$ LR_{BN}$&$ LR  $&$ LR_{CR}$&$ 
LR_{BN}$\\ \hline
1.2  &  9.17 &  9.07 &   9.06 &  2.06 &   2.06 &   2.07 \\
1.6  & 31.67 & 31.32 &  31.06 & 10.32 &  10.18 &   9.94 \\
2.0  & 58.54 & 58.05 &  57.43 & 24.56 &  24.21 &  23.50 \\
2.4  & 80.11 & 79.72 &  79.13 & 45.06 &  44.07 &  42.93 \\
2.8  & 91.05 & 90.93 &  90.49 & 62.15 &  61.08 &  59.56 \\
3.2  & 96.80 & 96.53 &  96.42 & 75.86 &  74.73 &  72.68 \\
3.6  & 99.05 & 99.01 &  98.85 & 85.86 &  84.74 &  83.10 \\
4.0  & 99.61 & 99.55 &  99.45 & 92.34 &  91.29 &  89.75 \\
\hline
\end{tabular}
\end{center}
\end{table}

Figure 3 plots the relative quantile discrepancies against the 
corresponding asymptotic quantiles, and Figure 4 plots the 
relative size distortions against the corresponding nominal 
levels of the tests when $\beta$ is the parameter 
of interest. Both figures clearly show that $LR_{CR}$ 
outperforms $LR$ and $LR_{BN}$. 

\begin{figure}[hp]
\label{fig3} 
\centering
\includegraphics[width=110mm,height=75mm]{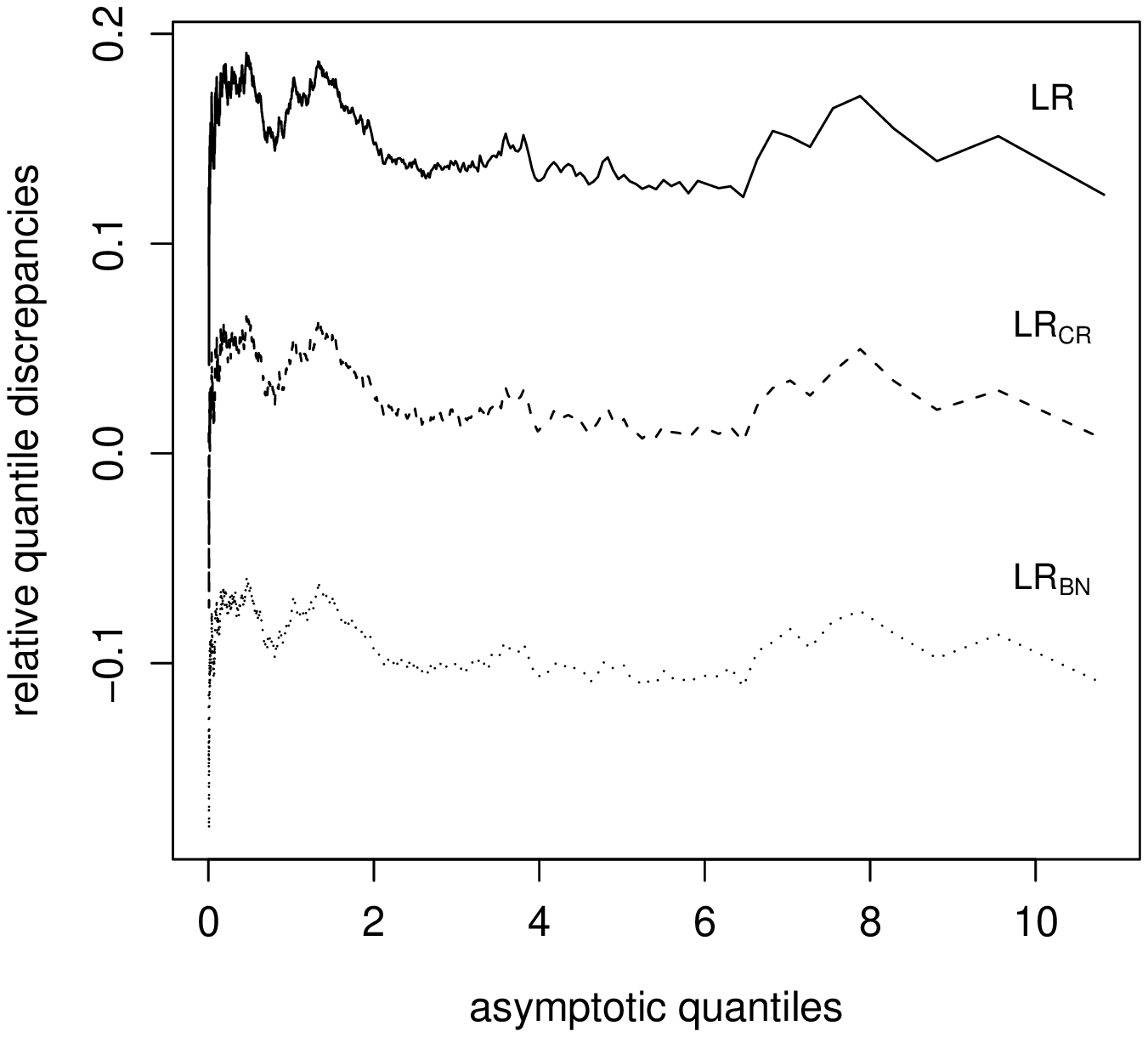}
\caption{Relative quantile discrepancy plot, inference on $\beta$.}
\includegraphics[width=110mm,height=75mm]{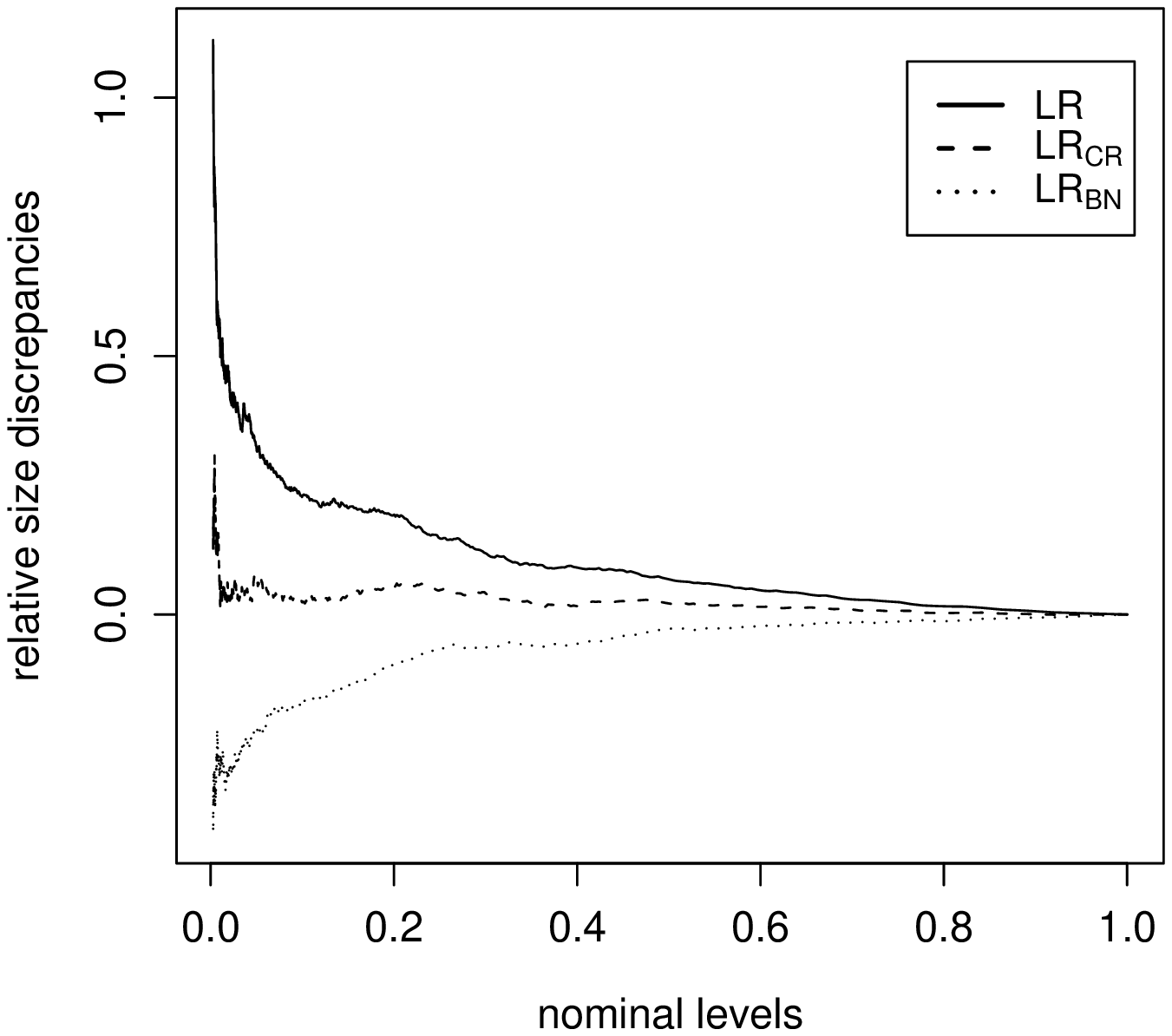}
\caption{Relative size distortion plot, inference on $\beta$.}
\end{figure}

\section{Additional numerical evidence: comparison with alternative 
estimators}\label{S:MonteCarlo2}

We shall now compare the small sample behavior of our adjusted profile 
maximum likelihood estimators of $\alpha$ to those proposed by Ng, 
Kundu and Balakrishnan~(2003) and From and Li~(2006), which are described 
in Section \ref{S:alternativeestimators}. The number of Monte Carlo 
replications is, as before, 10,000, the true values of $\alpha$ are 
0.5 and 1.0, and $n=10$. The simulation results are presented in Table 
\ref{T:alternativeestimators}. 

\begin{sidewaystable}\label{T:alternativeestimators}
\begin{center}
\caption{Point estimation of $\alpha$ revisited.}
\vspace{0.13cm}
\begin{tabular}{c c c c c c c c}\hline
\multicolumn{8}{c}{$\alpha=0.5$}\\ \hline
estimator               &mean   &bias       &variance  & MSE     &RB(\%)   &asymmetry&kurtosis\\ \hline
$\widehat\alpha_p$      &0.4625 & $-0.0375$ &   0.0121 &   0.0135&   7.5074&   1.3537&   5.9906\\
$\widehat\alpha_{CR}$   &0.4892 & $-0.0109$ &   0.0136 &   0.0138&   2.1695&   1.4258&   6.3098\\
$\widehat\alpha_{BN}$   &0.4893 & $-0.0107$ &   0.0137 &   0.0138&   2.1385&   1.4624&   6.3117\\ 
$\widetilde\alpha_{MME}$&0.4625 & $-0.0375$ &   0.0121 &   0.0135&   7.5077&   1.3537&   5.9906\\  
$\bar\alpha_{Ng}$       &0.4749 & $-0.0251$ &   0.0127 &   0.0133&   5.0275&   1.3874&   6.1376\\
$\bar\alpha_{Ng_{JMLE}}$&0.4573 & $-0.0427$ &   0.0133 &   0.0152&   8.5384&   1.3397&   5.9301\\                             
$\bar\alpha_{Ng_{JMME}}$&0.4579 & $-0.0421$ &   0.0133 &   0.0151&   8.4255&   1.3412&   5.9367\\                             
$\breve\alpha_{F1}$     &0.4625 & $-0.0375$ &   0.0121 &   0.0135&   7.4949&   1.3539&   5.9913\\                            
$\breve\alpha_{F2}$     &0.4690 & $-0.0310$ &   0.0166 &   0.0176&   6.2008&   1.3715&   6.0678\\                            
$\breve\alpha_{F3}$     &0.5065 & $\hfill 0.0065$ &   0.0326 &   0.0326&   1.3004&   1.4721&   6.5222\\                            
$\breve\alpha_{F4}$     &0.5033 & $\hfill 0.0033$ &   0.0220 &   0.0220&   0.6551&   1.4635&   6.4825\\ 
%\hline                          
\end{tabular}

\begin{tabular}{c c c c c c c c}\hline
\multicolumn{8}{c}{$\alpha=1.0$}\\ \hline
estimator               &mean  &bias        &variance  & MSE     &RB(\%)    &asymmetry&kurtosis\\ \hline
$\widehat\alpha_p$      &0.9160&  $-0.0840$ &   0.0472 &   0.0543&   8.4020&    2.3784& 11.8471 \\
$\widehat\alpha_{CR}$   &0.9752&  $-0.0248$ &   0.0548 &   0.0554&   2.4841&    2.4786& 12.5612 \\
$\widehat\alpha_{BN}$   &0.9792&  $-0.0208$ &   0.0567 &   0.0572&   2.0834&    2.4852& 12.6085 \\ 
$\widetilde\alpha_{MME}$&0.9158 & $-0.0842$ &   0.0471 &   0.0542&   8.4173&    2.3782& 11.8453 \\  
$\bar\alpha_{Ng}$       &0.9404 & $-0.0596$ &   0.0497 &   0.0533&   5.9615&    2.4207& 12.1452 \\
$\bar\alpha_{Ng_{JMLE}}$&0.9050 & $-0.0950$ &   0.0523 &   0.0613&   9.4959&    2.3591& 11.7120 \\                             
$\bar\alpha_{Ng_{JMME}}$&0.9061 & $-0.0939$ &   0.0522 &   0.0611&   9.3931&    2.3609& 11.7247 \\                             
$\breve\alpha_{F_1}$     &0.9168 & $-0.0832$ &   0.0475 &   0.0544&   8.3177&    2.3799& 11.8575 \\                            
$\breve\alpha_{F_2}$     &0.8916 & $-0.1084$ &   0.0763 &   0.0881&  10.8379&    2.3350& 11.5450 \\                            
$\breve\alpha_{F_3}$     &0.9974 & $-0.0026$ &   0.1279 &   0.1279&   0.2598&    2.5144& 12.8217 \\                            
$\breve\alpha_{F_4}$     &1.0039 & $\hfill 0.0039$ &   0.0906 &   0.0906&   0.3915&    2.5247& 12.8972 \\ \hline
\end{tabular}
\end{center}
\end{sidewaystable}

The figures in Table \ref{T:alternativeestimators} show that no estimator 
uniformly outperforms all others in terms of both bias and mean squared 
error. They also show that the adjusted profile maximum likelihood estimators 
are competitive, since they are amogst the best performing estimators in 
both situations ($\alpha = 0.5$ and $\alpha = 1.0$). When $\alpha = 0.5$, 
the least biased estimator is $\breve\alpha_{F4}$ (relative bias: 0.66\%)
whereas when the true parameter value is 1.0 the estimator 
$\breve\alpha_{F3}$ has the smallest relative bias (0.26\%). 
In both cases, $\widehat\alpha_{BN}$ is the estimator with the third 
smallest relative bias, followed by $\widehat\alpha_{CR}$. When $\alpha = 0.5$, 
for instance, the relative biases of these estimators are nearly four times 
smaller than those of the jackknife estimators and nearly 3.5 times smaller 
than the relative bias of the modified moments estimator. 

We note that the approach proposed in this paper, namely adjusting the profile 
log-likelihood function prior to maximization, has a clear advantage over 
the alternative approaches described in Section \ref{S:alternativeestimators}: 
it not only improves the small sample performance of point estimators, but 
also improves the finite sample behavior of associated likelihood ratio tests. 
That is, the correction delivers improved estimation \textit{and} testing 
inference in small samples. 

\section{Applications}\label{S:applications}

We shall now perform profile and adjusted profile likelihood inference 
using two real data sets. In both cases, we shall assume that observations 
are random draws from the Birnbaum--Saunders distribution. 

At the outset, we consider the data provided by Birnbaum--Saunders (1969b)
on the fatigue life of 6061-T6 aluminum coupons cut parallel to the 
direction of rolling and oscillated at 18 cycles per second (cps). 
The data set consists of 101 observations with maximum stress per cycle 
31,000 psi. Let $\alpha$ be the parameter of interest. The profile 
and adjusted profile maximum likelihood estimates are 
$\widehat\alpha=0.17038$, $\widehat\alpha_{CR}=0.17125$ and 
$\widehat\alpha_{BN}=0.17122$. Suppose we are interested in testing 
$\mathcal{H}_0:\alpha = 0.15$ against $\mathcal{H}_1:\alpha\not=0.15$. The test statistics 
based on $\ell_p(\alpha)$, $\ell_{CR}(\alpha)$ and $\ell_{BN}(\alpha)$
are, respectively, 3.5771, 3.8421 and 3.8351, with the following 
corresponding $p$-values: 0.05858, 0.04998 and 0.05019. Since the 
sample size is large (101 observations), the values of the three 
statistics are similar. However, the resulting inference is not the 
same at the 5\% nominal level, since the test based on  $\ell_{CR}(\alpha)$,
unlike the other two tests, yields rejection of the null hypothesis. 

We shall now turn to the case where $\beta$ is the parameter of 
interest. In particular, we are interested in testing $\mathcal{H}_0:\beta = 125$.
The test statistics are $LR=9.4279$, $LR_{CR}=9.3338$ and $LR_{BN}=9.2397$, 
with the following corresponding $p$-values: 0.00214, 0.00225 and 0.00237. 
Unlike the previous inference, here the three tests yield the same 
conclusion: the null hypothesis is rejected at the 1\% nominal level. 
The (profile and adjusted profile) maximum likelihood estimate of 
$\beta$ is 131.8188. 

The second application we consider uses data provided by McCool (1974).
The data describe the lifetime, in hours, of 10 sustainers of a certain 
type. They were used by Cohen, Whitten and Ding~(1984) as an illustration of 
the three-parameter Weibull distribution. The profile and adjusted 
profile maximum likelihood estimates of $\alpha$ are 
$\widehat\alpha=0.2825$, $\widehat\alpha_{CR}=0.2989$ and 
$\widehat\alpha_{BN}=0.2973$. Consider the test of 
$\mathcal{H}_0:\alpha=0.21$ against $\mathcal{H}_1:\alpha\not=0.21$. The test statistics 
are $LR=2.1646$, $LR_{CR}=2.8438$ and $LR_{BN}=2.7963$, with the 
following corresponding $p$-values: 0.1412, 0.0917 and 0.0945.
Therefore, the two adjusted profile likelihood ratio tests 
($LR_{CR}$ and $LR_{BN}$) reject the null hypothesis at the 10\% 
nominal level, unlike the likelihood ratio test ($LR$). Thus, the 
unadjusted and adjusted tests yield different conclusions. 

Now let $\beta$ be the parameter of interest. Its maximum likelihood 
estimate is $\widehat\beta=212.05$. Suppose we wish to test 
$\mathcal{H}_0:\beta=180$ against $\mathcal{H}_1:\beta\not=180$. The test statistics 
are $LR=2.9417$, $LR_{CR}=2.6415$ and $LR_{BN}=2.3414$; 
the respective $p$-values are 0.0863, 0.1041 and 0.1260.
Again, the use of adjustments to the profile likelihood function 
makes a difference: unlike the usual likelihood ratio test, 
the adjusted likelihood tests reject the null hypothesis at the 
10\% nominal level.  

\section{Concluding remarks}\label{S:conclusions}

We considered the issue of performing inference on the parameters 
that index the Birnbaum--Saunders distribution. More specifically, 
we have focused on the situation where one wishes to make inference 
on one of the parameters, the other parameter being of nuisance fashion. 
Using the results in Cox and Reid (1987) and in 
Barndorff--Nielsen (1983), we derived two adjustments that can applied 
to the profile likelihood function so as to deliver improved inference. 
Approximations due to  Severini (1998, 1999) were used in order to 
obtain one of such adjustments. Monte Carlo simulation results  
have shown that the adjusted estimators and tests --- i.e., estimators 
and tests based on the adjusted profile likelihood functions --- can 
deliver more accurate inference than that carried out using 
the usual maximum likelihood estimator and the standard likelihood 
ratio test in small samples. 
In particular, the adjusted estimators displayed smaller 
biases and the adjusted tests, smaller size distortions. 
For instance, we reported Monte Carlo simulation results in which the 
usual likelihood ratio test displayed null rejection rate of nearly 
9.5\% at the 5\% nominal level whereas the sizes of our two adjusted 
tests where 5.2\% and 4.6\%, and in which the relative bias of our two 
estimators
were approximately four times smaller than that of the maximum likelihood 
estimator. The adjusted profile likelihood tests have also outperformed 
the Bartlett-corrected likelihood ratio test. 
We recommend the use of the adjusted profile likelihood inference 
developed in this paper to practitioners who wish to model 
reliability data using the Birnbaum--Saunders model. In particular, 
we recommend the use of the Cox--Reid adjusted profile likelihood
function, since it yielded the most reliable (hypothesis testing) 
inference when the parameter 
was of interest was $\beta$ and it was competitive with the inference 
obtained using the Barndorff--Nielsen modified profile likelihood function
when $\alpha$ was the parameter of interest.  
In future research, 
we shall obtain adjustments to Birnbaum--Saunders profile likelihoods under 
data censoring. 

\section*{Acknowledgements}

\noindent Research supported by CNPq and CAPES. We also thank two 
anonymous referees for their constructive comments and suggestions.

\end{document}